\documentclass[prl,singlecolumn,showpacs,preprintnumbers,amsmath,amssymb]{revtex4}

\usepackage{mathrsfs}
\usepackage{graphicx}
\usepackage{dcolumn}
\usepackage{bm}

\begin{document}

\title{Stochastic Energetics for Non-Gaussian Processes}

\author{Kiyoshi Kanazawa$^1$, Takahiro Sagawa$^{1,2}$, and Hisao Hayakawa$^1$}

\affiliation{$^1$Yukawa Institute for Theoretical Physics, Kyoto University, Kitashirakawa-oiwake cho, Sakyo-ku, Kyoto 606-8502, Japan \\
	  $^2$The Hakubi Center, Kyoto University, Yoshida-ushinomiya cho, Sakyo-ku, Kyoto 606-8302, Japan}
\date{\today}

\begin{abstract}
By introducing a new stochastic integral, we investigate the energetics of classical stochastic systems driven by non-Gaussian white noises.
In particular, we introduce a decomposition of the total-energy difference into the work and the heat for each trajectory,
and derive a formula to calculate the heat from experimental data on the dynamics.
We apply our formulation and results to a Langevin system driven by a Poisson noise.
\end{abstract}


\maketitle

\section{Introduction}
	%
	Stochastic processes driven by non-Gaussian noises have been shown powerful to analyze various natural and social phenomena; 
	the shot noise in electrical circuits~\cite{Blanter}, L\'evy flights of bumblebees~\cite{Bee} and human mobilities,~\cite{Gonzalez}, and random walks in finance and econophysics~\cite{Cont,Mantegna}. 
	Moreover, non-Gaussian effects have been discussed for small thermodynamics systems such as biological molecular machines.
	For example, a non-Gaussian white noise has been used for modeling directed transport in Brownian motors~\cite{Reimann,Luczka},
	and the Adenosine-triphosphate (ATP) reception by red-blood-cell membranes~\cite{Gov,RedBloodCell} has been modeled in terms of non-Gaussian noises. 
	This Letter aims to introduce a new stochastic integral for non-Gaussian processes and, to apply it to small thermodynamic systems.

	Recent progress in experimental technique causes growing interest in small thermodynamic systems~\cite{Carlos}.
	In such systems, thermodynamic quantities become stochastic due to
	environmental fluctuations that can be non-Gaussian.
	The thermodynamic energy balance in a single trajectory has been formulated by stochastic energetics~\cite{Sekimoto1},
	in which thermodynamic quantities such as work and heat are defined for each trajectory~\cite{Sekimoto2}.
	Stochastic energetics has been widely applied to theories~\cite{Seifert1,Jarzynski,Sagawa,Kawai,Evans,Kurchan,Crooks,Maes,Harada,Hatano}
	and experiments~\cite{Seifert2,Liphardt,Toyabe1,Collin,Toyabe2,Trepagnier} of modern nonequilibrium statistical mechanics.

	For Gaussian processes, it has been established that the Stratonovich-type stochastic calculus is consistent with 
	the stochastic energetics~\cite{Sekimoto1}.
	In fact, the Stratonovich calculus has been used for experimental and numerical verifications of nonequilibrium equalities 
	such as the fluctuation theorem~\cite{Seifert2,Liphardt,Toyabe1,Collin,Toyabe2,Trepagnier}.
	This is because the ordinary differential calculus, such as the chain rule
	$df/dt=(df/dx)(dx/dt)$, is satisfied for the Stratonovich stochastic calculus.
	In contrast, the It\^o-type stochastic calculus needs an alternative formula, which is known as the It\^o rule,
	instead of the ordinary chain rule.
	So far, however, stochastic energetics for non-Gaussian processes has not been fully investigated.
	It is remarkable that, as will be shown later, the Stratonovich calculus 
	is inadequate to formulate the stochastic energetics for non-Gaussian processes. 

	In this Letter, we introduce a new stochastic integral, with which the ordinary differential calculus is applicable to non-Gaussian processes.
	Based on it, we investigate the decomposition of the total-energy difference into the work and the heat for each trajectory,	
	and derive a formula that is applicable to measurements of the heat in experiments.
	We show that our stochastic energetics is consistent with the first law of thermodynamics.	
	
\section{Stochastic integrals}
	Let $\hat \xi(t)$ be a white noise with $\langle \hat \xi (t) \rangle = 0$ and
	$\langle \hat \xi(t)\hat \xi(s) \rangle = \delta (t-s)$,
	where the bracket denotes the ensemble average of a stochastic variable.
	We consider a stochastic differential equation
	$d\hat X(t)/dt=a(\hat X(t))+b(\hat X(t))\hat \xi(t)$,
	where $\hat X(t)$ is the phase-space point of a Brownian particle, 
	and $a(\hat X(t)), b(\hat X(t))$ are arbitrary functions of $\hat X(t)$.
	The stochastic differential equation can be rewritten as the integral form
	$\hat X(t) = \hat X_0 + \int_0^t ds a(\hat X(s))+\int_0^t ds\hat \xi(s)b(\hat X(s))$,
	where the last term on the right-hand side (rhs) involves a stochastic integral.
	In usual stochastic calculus for Gaussian processes, we use one of the following two definitions:
	\begin{align}\label{Ito}
		\quad \int_{0}^{t}ds\hat \xi(t)\cdot b(\hat X(t)) 	&\equiv \lim_{\Delta t \rightarrow +0}
												\sum_{i=1}^n\Delta t \hat \xi(t_i)b(\hat X_i),\\
		\quad \int_{0}^{t}ds\hat \xi(t)\circ b(\hat X(t)) 	&\equiv \lim_{\Delta t \rightarrow +0}
												\sum_{i=1}^n\Delta t \hat \xi(t_i)b\left( \frac{\hat X_i+\hat X_{i+1}}{2}\right) ,
	\label{Stratonovich}
	\end{align}
	where the symbols $\cdot$ and $\circ$ denote the It\^o calculus and the Stratonovich one, respectively,
	and $\Delta t\equiv t/n, t_i\equiv i\Delta t, \hat X_i\equiv \hat X(t_i).$
	The term $\Delta t\hat\xi(t_i)$ is shorthand for $\Delta \hat W(t_i) \equiv \hat W(t_{i+1})-\hat W(t_i)$, 
	where $\hat W(t) \equiv \int_{0}^{t}ds \hat \xi(s)$ is the Wiener process. 

	Our strategy to define the new stochastic integral for non-Gaussian processes is
	the white noise limit of a non-Gaussian colored noise.
	First, we construct the colored noise by using white noises.
	Let $\epsilon$ be a time constant to characterize the correlation time of the noise,
	and let $\Delta t\hat \xi(t_i)$ be shorthand for $\Delta \hat L(t_i)\equiv\hat L(t_{i+1})-\hat L(t_i)$, 
	where $\hat L(t)\equiv\int_0^t ds\hat \xi(s)$ is the L\'evy process.
	We define the colored noise:
	\begin{equation}\label{eq:def_integral}
		\Delta t\hat \xi_{\epsilon}(t_i) \equiv \frac{1}{l}\sum_{j=0}^{l-1}\Delta t\hat \xi (t_{i-j}), 
	\end{equation}
	where $l \equiv \epsilon / \Delta t$ is assumed to be an integer.
	The correlation of the noise satisfies $\left<\hat \xi_\epsilon(t) \hat \xi_\epsilon(s) \right>= 0$ only if $|t-s|\geq \epsilon$.
	We call this manipulation $\epsilon$-smoothing.
	For example, the white Poisson noise $\hat\xi(t)=\sum_{i}\delta (t-\hat t_i)$, 
	where each Poisson flight happens at $\hat{t}_i$, is used to construct the colored noise as shown in Fig.~1.
	\begin{figure}
		\centering
		\includegraphics[width=80mm]{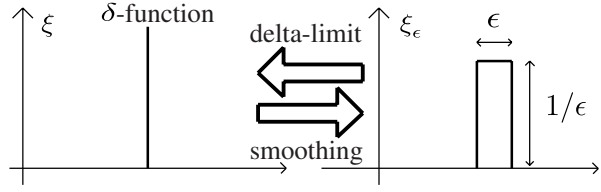}
		\caption{A schematic of the $\epsilon$-smoothing. }
		\label{fig:smoothing}
	\end{figure}
		
	Introducing a new symbol $\ast$, we propose the following stochastic integral for non-Gaussian processes as a limit of the colored noise:
	\begin{equation}\label{eq:4}
		\int_{0}^{t}ds\hat \xi(t)\ast f(\hat X(t)) 	\equiv \lim_{\epsilon \rightarrow +0}\lim_{\Delta t \rightarrow +0}
													\sum_{i=1}^n\Delta t\hat \xi_{\epsilon}(t_i)f(\hat X(t_i)),
	\end{equation}
	where $f(x)$ is an arbitrary function.
	We refer to this integral as the $\ast$-integral or the $\ast$-calculus.
	Correspondingly, we interpret the stochastic differential equation with the $\ast$-integral as 
	\begin{equation}\label{eq:SDE}
		\frac{d\hat X(t)}{dt}=a(\hat X(t))+b(\hat X(t))\ast \xi(t).
	\end{equation}
	The two limits in Eq. (\ref{eq:4}) are not commutable.
	In the It\^o integral, the limits are taken simultaneously $\Delta t=\epsilon \rightarrow +0$.
	On contrary, the limit of $\Delta t \rightarrow +0$ is taken before the limit of $\epsilon \rightarrow +0$ in our formulation.
	We note that our integral is equivalent to the Stratonovich one for Gaussian processes~\cite{WongZakai}.

	Next, let us discuss the transformation formula from the $\ast$-integral to the It\^o one. 
	By using H\"anggi's functional formula~\cite{Novikov,Hanggi}, we obtain (see supplementary material for a derivation)
	\begin{equation}
		d\hat L(t)\ast f(\hat X(t))\!=\!\sum_{n=0}^{\infty}\!\frac{d\hat L^{n+1}(t)}{(n+1)}\cdot\left( \! \left\{ b(x)\! \displaystyle \frac{\partial}{\partial x}\right \}^n \! f(x)\! \Bigg|_{x=\hat X(t)}\!\right),
						\label{eq:transform_formula}
	\end{equation}
	where $b(\hat X(t))$ describes the same function in Eq.~(\ref{eq:SDE}), and $f(\hat X(t))$ is an arbitrary function of $\hat X(t)$.
	For Gaussian processes, Eq.~(\ref{eq:transform_formula}) reduces to the transformation formula 
	from the Stratonovich integral to the It\^o one, because $d\hat L^n = 0$ holds for $n \geq 3$.
	Although other types of $\epsilon$-smoothing can be used to construct the colored noise,
	Eq.~(\ref{eq:transform_formula}) does not depend on the way of the $\epsilon$-smoothing~\cite{Kanazawa}.
	We note that Eq.~(\ref{eq:transform_formula}) can be regarded
	as a straightforward generalization of the result in Refs.~\cite{Paola,Proppe}, 
	in which Eq.~(\ref{eq:SDE}) has been transformed into the It\^o type.
	
\begin{figure}
		\centering
		\includegraphics[width=55mm]{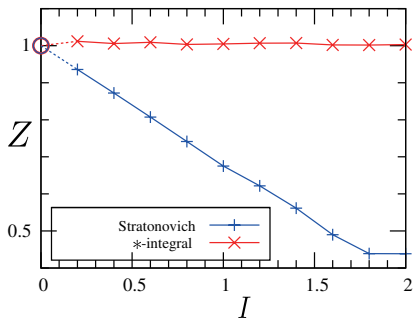}
		\caption{(Color online) Numerical results on the consistency between Eq. (\ref{eq:BlackScholes}) and $\hat Y$.
		We fix the variance of the Poisson noise $\lambda I^2=4.0$ and simulate the equation until $t=1.0$ with $\epsilon=0.001$.
		As $I$ increases, the difference betweent the $\ast$- and the Stratonovich integrals becomes significant.
		In the Gaussian limit of $I\rightarrow 0$, 
		$Z$ is expected to equal unity in the both calculuses as denoted by an open circle.}
		\label{fig:ordinary_calculus}
	\end{figure}

	As a simple example, we apply the $\ast$-calculus to the Black-Scholes equation driven by a Poisson noise~\cite{Gardiner}
	\begin{equation}\label{eq:BlackScholes}
		\frac{d\hat X}{dt}=(-1 + \hat \xi)\hat X,
	\end{equation}
	where $\hat \xi=\hat \eta -\lambda I$ with a Poisson noise $\hat\eta$ characterized by its intensity $I$ and transition rate $\lambda$.
	The Poisson noise is a typical non-Gaussian noise.
	In fact, any noise can be decomposed into the combination of Gaussian and Poisson noises~\cite{Ito}.
	If we define $\hat \xi \hat X$ on the rhs of Eq. (\ref{eq:BlackScholes})
	based on the $\ast$-integral as $\hat \xi \ast \hat X$,
	then the solution of Eq. (\ref{eq:BlackScholes}) equals the following quantity:
		$\hat Y= \exp{\left[-t+\int_0^t ds\hat \xi(s) \right]},$
	which is the formal solution of the Black-Scholes equation with the ordinary differential calculus.
	In contrast, if we adopt the Stratonovich integral to define the rhs of Eq.~(\ref{eq:BlackScholes}), 
	its solution does not equal $\hat Y$.
	Figure~\ref{fig:ordinary_calculus} shows the numerical results on $Z\equiv \left <\hat Y/\hat X\right >$
	for the cases of the $\ast$-integral and the Stratonovich one.
	The $\epsilon$-smoothing is performed by the definition Eq.~(\ref{eq:def_integral}).
	As $I$ increases, $Z$ becomes different from 1 for the Stratonovich integral due to the non-Gaussian property of the noise.
	In contrast, the $\ast$-calculus keeps $Z\approx 1$ independently of the value of $I$.
	We also note that the result of the It\^o calculus is ranged around $Z=5$.
	These results imply that neither It\^o nor Stratonovich calculus is obtained 
	as the white noise limit of a colored noise.

\section{Stochastic Energetics for Non-Gaussian Processes}
	We now apply the $\ast$-integral to formulate stochastic energetics for non-Gaussian processes.
	Let us consider the following underdamped Langevin equations
	\begin{equation}
		\frac{d\hat p}{dt}= -\frac{\gamma}{m}\hat p -\frac{\partial U(\alpha,x)}{\partial x}\Bigg|_{x=\hat x} +g(\hat x,\hat p)\ast \hat \xi,
		\quad	\frac{d\hat x}{dt}= \frac{\hat p}{m},
	\end{equation}
	where $\hat \xi$ is a non-Gaussian noise whose mean value is zero,
	$\hat x$ is the position of the particle, $\hat p$ is its momentum, 
	and $U(\alpha,x)$ is the potential with an external parameter $\alpha$, such as the intensity of optical tweezers.
	If $g(x,p)$ is constant, the noise is called additive. Otherwise, it is called multiplicative. 
	It is known that Brownian motion near a wall can be characterized by a multiplicative noise~\cite{Lau}.
	The detailed balance condition is not assumed in this Letter, and the following results can be applied to athermal systems. 
	We note that the detailed balance condition in athermal systems has been studied in Ref.~\cite{Hanggi2}.

	Let us first define the thermodynamic quantities for each trajectory.
	We divide the total-energy difference into the following two parts
	\begin{equation}
		d\hat W =\frac{\partial U}{\partial \alpha}\Bigg|_{x=\hat x}d\alpha,
		\quad d\hat Q = \left(-\frac{\gamma \hat p^2}{m^2} + \frac{\hat \xi\ast g(\hat x,\hat p)\hat p}{m}\right )dt,\label{eq:def_Q}
	\end{equation}
	where $\hat W$ is the mechanical work through the parameter $\alpha$,
	and $\hat Q$ is the energy flow induced by the microscopic degrees of freedom in the environment.
	The usual heat is included in $\hat Q$.
	Our formalism has practical utilities for experimental and numerical data analysis of the heat.
	We call $\hat Q$ as ``heat" for convenience.
	By using the ordinary chain rule for the $\ast$-integral, we confirm the first law of thermodynamics:
	\begin{equation}
		d\hat E	 = \frac{\hat p\ast d\hat p}{m} + \frac{\partial U}{\partial x}\ast d\hat x + \frac{\partial U}{\partial \alpha} d\alpha
				 = d\hat Q + d\hat W,
	\end{equation}
	where $\hat E \equiv \hat p^2/2m+U(\alpha,\hat x)$ is the internal energy of the particle.

	We next derive a representation of the average heat flux by using the $\ast$-calculus.
	For this purpose, we first write down the Kramers equation for $P(x,p,t) = \left <\hat {\mathcal{P}}(x,p,t)\right >$
	with $\hat {\mathcal{P}}(x,p,t)=\delta(x-\hat x(t))\delta(p-\hat p(t) )$. 
	The stochastic Liouville equation~\cite{Kubo} is given by
	\begin{equation}
		\frac{\partial}{\partial t}\hat {\mathcal{P}}(x,p,t)	+ \frac{\partial}{\partial x}\{ \dot{\hat x}(t) \ast\hat {\mathcal{P}}(x,p,t) \} \notag
													+ \frac{\partial}{\partial p}\{ \dot{\hat p}(t) \ast\hat {\mathcal{P}}(x,p,t) \} = 0,\notag
	\end{equation}
	where the stochastic partial differential equation is regarded as the $\ast$-calculus.
	Taking average of this equation and using the transformation formula~(\ref{eq:transform_formula}),
	we obtain the generalized Kramers equation
	\begin{equation}
		\frac{\partial}{\partial t}P   = -\frac{\partial}{\partial x}\left( \frac{p}{m}P\right) 
		-\frac{\partial}{\partial p}\left [\left(-\frac{\gamma p}{m} - \frac{\partial U}{\partial x}\right)P
		+\left <\hat \xi \ast g\hat {\mathcal{P}}\right >\right ],       \label{eq:generalized_Kramers}
	\end{equation}
	where $\left <\hat \xi \ast g\hat {\mathcal{P}}\right >$ describes the diffusion of the particle.
	Transforming this diffusion term into the It\^o type, we reproduce the ordinary Kramers equation.
	By introducing
	\begin{equation}
		J_x	\equiv	 \frac{p}{m}P, \>\>
		J_p	\equiv	 \left( -\frac{\gamma p}{m}-\frac{\partial U}{\partial x}\right) P + \left <\hat \xi \ast g\hat {\mathcal{P}}\right >,	\label{J_p}
	\end{equation}
	we rewrite (\ref{eq:generalized_Kramers}) as the conservation of the probability
	$\partial P/\partial t+\partial J_x/\partial x+\partial J_p/\partial p=0.$
	Note that the probability fluxes include infinite number of differentiations if we use the It\^o calculus.
	The probability-conservation formula leads to a simple 
	identity of the total derivative for any function $f(x,p,\alpha)$:
	\begin{equation}
		\left <\frac{d}{dt}f(\hat x,\hat p,\alpha)	\right>= \int dxdp
												\left(
													  \frac{\partial f}{\partial x}J_x+\frac{\partial f}{\partial p}J_p
												\right)
											+ \dot \alpha\left <\frac{\partial f}{\partial \alpha}\right >.
	\end{equation}
	Taking $f(x,p,\alpha)$ as $E(x,p,\alpha)$,
	we derive that the heat satisfies
	formula of the average of heat flux
	\begin{equation}\label{eq:Avg_Heat_Flux}
		\left <\frac{d \hat Q}{dt}\right >=\int dxdp\left( \frac{\partial E}{\partial x}J_x+\frac{\partial E}{\partial p}J_p\right).
	\end{equation}
	We note that we can straightforwardly apply the $\ast$-calculus to the overdamped Langevin equation.

\section{Heat Measurement Formula}
	In the case of the underdamped Langevin equation driven by an additive non-Gaussian noise,
	the transformation formula Eq. (\ref{eq:transform_formula}) can be reduced to 
	\begin{equation}\label{eq:Heat_Measurement}
		d\hat Q = -\frac{\gamma\hat p^2}{m^2}dt + d\hat L\ast  \frac{\hat p}{m}
				= -\frac{\gamma\hat p^2}{m^2}dt + d\hat L\cdot \frac{\hat p}{m}  + \frac{(d\hat L)^2}{2}.
	\end{equation}
	This formula can be applied to measurements of the heat in experiments,
	because Eq. (\ref{eq:transform_formula}) includes the noise term up to only the second order.
	The term $(d\hat L)^2$ is not deterministic but stochastic, 
	contrary to the cases of the Gaussian noise,
	in which $(d\hat L)^2$ can be replaced by $\sigma^2dt$ with the variance $\sigma$.
	We note that the formula for the heat measurement becomes more complicated for multiplicative noises.
	We also note that the heat measurement formula for the overdamped Langevin equation is not simple even for additive noises, 
	where higher order cumulants appear for the cases of a non-harmonic potential~\cite{Kanazawa}.

\section{A Model of ATP reception}
	To demonstrate how a non-Gaussian feature appears in a stochastic process,
	let us analyze a Brownian particle in an ATP bath.
	When the particle receive an ATP, the particle suddenly moves.		
	We assume that the particle obeys the overdamped Langevin equation under a non-harmonic potential
	$U(x) = (k/2)x^2+(\epsilon/4)x^4$ with small $\epsilon$.
	The system is driven by an additive Poisson noise whose transition rate $P(x\rightarrow y)$ is given by
	$P(x \rightarrow y)=		\lambda/2$ (if $\quad  x-y=\pm I/\gamma$),
								$0$ (otherwise),
	where $I$ is the intensity of the Poisson noise and $\gamma$ is a friction constant.
	We note that the detailed balance condition is violated in this model.
	The dimensionless Langevin equation of this system is given by
	\begin{align}\label{ATP_eq}
		\frac{d\hat X}{d\tau} 		&=-\hat X-\tilde{\epsilon}\hat X^3+\hat {\xi},\>\>\> d\hat{L}=\hat{\xi}d\tau,\\
		\left <d\hat {L}^n\right >	&=	\displaystyle\begin{cases}
											0 & ( n$: odd$ ) ,\cr
											\left( t_p/t_s \right)^{n/2-1}d\tau & ( n$: even$ ),
										\end{cases}
	\end{align}
	where we introduced the characteristic constants and dimensionless parameters as
	$
		t_s= \gamma/k, 
		t_p=1/\lambda,
		x_s= (I/\gamma)\sqrt{(t_s/t_p)},
		\tilde{\epsilon}=x_s^2\epsilon/k,
		\hat x= x_s\hat X,
		$ and $
		t=t_s\tau.
	$
	Note that $t_s$ and $t_p$ are the characteristic time scales of the relaxation of the system and the Poisson noise, respectively, 
	and therefore $t_p/t_s$ characterizes non-Gaussian effects.
	By analyzing this model, we demonstrate that
	$\ast$-integral is consistent with the first law of thermodynamics
	and that the non-Gaussian effects are relevant for $t_p \sim t_s$.

	We first demonstrate that the first law of thermodynamics is consistent with the $\ast$-calculus in this system.
	Let us define the three types of dimensionless energies
	$
		\tilde{U}		\equiv \frac{\hat X^2}{2} + \frac{\epsilon\hat X^4}{4},
		\tilde{U}^* 	\equiv \int_0^\tau d\hat X(s) \ast  \frac{\partial \tilde{U}}{\partial x},
		\tilde{U}^S 	\equiv \int_0^\tau d\hat X(s) \circ \frac{\partial \tilde{U}}{\partial x},
	$
	where $\tilde U$ is the total potential energy, 
	$\tilde U^\ast$ is the heat with the $\ast$-calculus, 
	and $\tilde U^S$ is that with the Stratonovich calculus.
	We note that the work is zero in this case.
	Figure 3 shows the time evolution of the three quantities without taking the ensemble average,
	where $\tilde{U}^*$ is consistent with the first law of thermodynamics
	while $\tilde{U}^S$ is not.
	\begin{figure}
		\includegraphics[width=60mm]{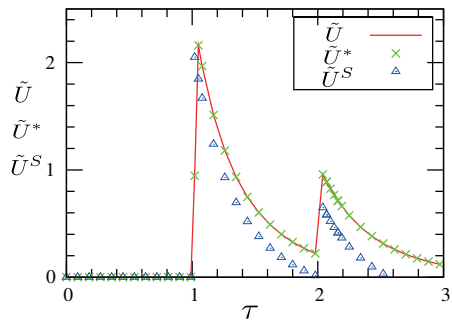}
		\caption{
		(Color online) 
		The check of the first law of thermodynamics through the numerical solution of Eq. (\ref{ATP_eq})
		with the intensity $I=\pm 2.0$, where the Poisson noises are added at $\tau=1.0$ and $\tau=2.0$.
		The result of $\tilde{U}^*$ based on the $\ast$-calculus agrees with the total energy $\tilde{U}$, 
		while $\tilde{U}^S$ based on the Stratonovich calculus is different from $\tilde{U}$.
	}
	\end{figure}
	
	We next derive the condition where non-Gaussian effects are relevant.
	The average of the energy in the steady state can be obtained for small $\epsilon$ as
	$
		\lim_{\tau \to \infty} \left <\tilde{U}\right > = 
			1/4-3\tilde{\epsilon}/16-\left(\tilde{\epsilon}/16\right)\left (t_p/t_s\right).
	$
	This implies that we cannot ignore the non-Gaussian effect for $t_p\approx t_s$, while
	the non-Gaussian effect vanishes for $t_s \gg t_p$.
	In fact, $t_p$ is the relaxation time
	in which the Poisson noise converges to a Gaussian noise according to the central limit theorem.
	For $t_p \approx t_s$, the system evolves before the relaxation, and 
	therefore we cannot replace the Poisson noise by a Gaussian noise.
	In contrast, we can adiabatically eliminate the non-Gaussian effect in the Poisson process for $t_s \gg t_p$.

\section{Conclusion}
	In this Letter, we have developed the stochastic energetics of small thermodynamic systems driven by a non-Gaussian noise,
	by introducing the new stochastic integral~(\ref{eq:4}) which we refer to as the $\ast$-integral.
	The investigation of the second law of thermodynamics and fluctuation theorem
	for non-Gaussian processes is a future issue, in which the $\ast$-calculus would play an important role.

	
	
\section{Acknowledgments}
\begin{acknowledgments}
	We are grateful to H. Nakao and S.-I. Sasa 
	for valuable discussions.
	This work was supported by the Global COE Program 
	``The Next Generation of Physics, Spun from Universality and Emergence"
	from the Ministry of Education, Culture, Sports, Science and Technology (MEXT) of Japan,
	the Grant-in-Aid of MEXT (Grants No. 21540384), and 
	the Grant-in-Aid for Research Activity Start-up (Grants No. 11025807).

\end{acknowledgments}

\section{Supplementary Information: The transformation formula from the $\ast$-integral to the It\^o integral}
		We consider the following Langevin equation
		\begin{equation}
			\frac{d\hat X(t)}{dt}=a(\hat X(t))+b(\hat X(t))\ast \hat \xi(t),
		\end{equation}
		where $a(\hat X(t)), b(\hat X(t))$ are arbitrary functions of $\hat X(t)$ and $\hat \xi(t)$ is a non-Gaussian white noise.
		Let $f(\hat X(t))$ be an arbitrary function of $\hat X(t)$ and let $\hat L(t) \equiv \int_0^tds\hat \xi(t)$ be the L\'evy process. 
		Here we prove Eq. (6) of the main text.
		
		We have formulated the $\ast$-calculus as the white noise limit of a colored noise.
		We use H\"anggi's functional formula~\cite{Novikov,Hanggi} for stochastic processes driven by a colored noise,
		and obtain
		\begin{equation}
			\left<\hat \eta(t) g[\hat \eta]\right>=\sum_{n=1}^\infty \frac{1}{n!}\int_0^t ds_1\dots ds_n C_{n+1}(t,s_1,\dots,s_n)\left<\frac{\delta^n g[\hat \eta]}{\delta \hat \eta(s_1)\dots\delta \hat \eta(s_n)}\right>,
		\end{equation}
		where $\hat \eta$ is a colored noise, $g[\hat \eta]$ is an any functional of $\hat \eta$, 
		and $C_{n+1}(t,s_1,\dots,s_n)$ is a $(n+1)$-point cumulant function defined by
		\begin{equation}
			C_{n+1}(t,s_1,\dots ,s_n) \equiv \frac{i^{-(n+1)}\delta^{n+1}}{\delta v(s)\delta v(s_1)\dots \delta v(s_n)}\left[\frac{1}{v(t)}\frac{\partial}{\partial t}\Phi [v]\right],\>\>\>\>
			\Phi[v] \equiv \log\left[\left<\exp{\left(i\int_0^t ds v(s)\hat \eta(s)\right)}\right>\right].
		\end{equation}
		We note that $f(\hat X(t))$ is a functional of $\hat \eta$ because $\hat X(t)$ is a functional of $\hat \eta$.
		We substitute $\hat \eta$ and $g[\hat \eta]$ with $\hat \xi_\epsilon$ and $f(\hat X(t))$ respectively, and take the limit of $\epsilon \rightarrow +0$.
		We then obtain
		\begin{equation}\label{eq:1}
			\left<\hat \xi(t) \ast f(\hat X(t))\right>	=\lim_{\epsilon \rightarrow +0}\left<\hat \xi_{\epsilon}(t)f(\hat X(t))\right>
			=\sum_{n=0}^\infty \frac{C_{n+1}}{(n+1)!}\left<\left(b(x)\frac{\partial}{\partial x}\right)^n f(x)\Bigg|_{x=\hat X(t)}\right>,\\
		\end{equation}
		where $C_{n}=\left<d\hat L^n(t)/dt\right>$,
		or equivalently,
		\begin{equation}\label{eq:2}
			\left<d\hat L(t)  \ast f(\hat X(t))\right>	=\sum_{n=0}^\infty \frac{\left<d\hat L^{n+1}(t)\right>}{(n+1)!}\left<\left(b(x)\frac{\partial}{\partial x}\right)^n f(x)\Bigg|_{x=\hat X(t)}\right>.
		\end{equation}
		It is remarkable that the averages of $d\hat L^n(t)$ and $\hat X(t)$ in the right-hand sides (rhs') of Eqs.~{(\ref{eq:1})} and {(\ref{eq:2})} are decoupled. 
		This is known as the non-anticipating property, which is an intrinsic character of the It\^o integral~\cite{Gardiner}.
		Therefore, the rhs of Eq.~{(\ref{eq:2})} can be written in the It\^o form as
		\begin{equation}
			\left<d\hat L(t)  \ast f(\hat X(t))\right>	=\sum_{n=0}^\infty \left<\frac{d\hat L^{n+1}(t)}{(n+1)!}\cdot \left(b(x)\frac{\partial}{\partial x}\right)^n f(x)\Bigg|_{x=\hat X(t)}\right>.
		\end{equation}
		We can remove the averaging operator $\left<\dots\right>$ because this equation holds for an arbitrary function $f(x)$,
		and obtain 
		\begin{equation}
			d\hat L(t) \ast f(\hat X(t)) 	= \sum_{n=0}^\infty \frac{d\hat L^n(t)}{(n+1)!} \cdot \left(\left\{ b(x)\frac{\partial}{\partial x}\right\}^{n-1}f(x)\right) \Bigg|_{x=\hat X(t)}
											=\frac{e^{d\hat L(t)\cdot b(x)\frac{\partial}{\partial x}}-1}{b(x)\frac{\partial}{\partial x}}\cdot f(x)\Bigg|_{x=\hat X(t)},
		\end{equation}	
		which implies Eq. (6) of the main text.

\end{document}